\documentclass[dvips]{article}
\usepackage{supertabular,lscape,epsfig}
\usepackage{amssymb}
\usepackage{amsmath}

\DeclareSymbolFont{ppa}{OT1}{ppl}{m}{it}
\DeclareMathSymbol{\vv}{\mathalpha}{ppa}{'166}

\begin{document}

\newcommand{\dd}{\,{\rm d}}
\newcommand{\ie}{{\it i.e.},\,}
\newcommand{\etal}{{\it et al.\ }}
\newcommand{\eg}{{\it e.g.},\,}
\newcommand{\cf}{{\it cf.\ }}
\newcommand{\vs}{{\it vs.\ }}
\newcommand{\zdot}{\makebox[0pt][l]{.}}
\newcommand{\up}[1]{\ifmmode^{\rm #1}\else$^{\rm #1}$\fi}
\newcommand{\dn}[1]{\ifmmode_{\rm #1}\else$_{\rm #1}$\fi}
\newcommand{\upd}{\up{d}}
\newcommand{\uph}{\up{h}}
\newcommand{\upm}{\up{m}}
\newcommand{\ups}{\up{s}}
\newcommand{\arcd}{\ifmmode^{\circ}\else$^{\circ}$\fi}
\newcommand{\arcm}{\ifmmode{'}\else$'$\fi}
\newcommand{\arcs}{\ifmmode{''}\else$''$\fi}
\newcommand{\MS}{{\rm M}\ifmmode_{\odot}\else$_{\odot}$\fi}
\newcommand{\RS}{{\rm R}\ifmmode_{\odot}\else$_{\odot}$\fi}
\newcommand{\LS}{{\rm L}\ifmmode_{\odot}\else$_{\odot}$\fi}

\newcommand{\Abstract}[2]{{\footnotesize\begin{center}ABSTRACT\end{center}
\vspace{1mm}\par#1\par
\noindent
{~}{\it #2}}}

\newcommand{\TabCap}[2]{\begin{center}\parbox[t]{#1}{\begin{center}
  \small {\spaceskip 2pt plus 1pt minus 1pt T a b l e}
  \refstepcounter{table}\thetable \\[2mm]
  \footnotesize #2 \end{center}}\end{center}}

\newcommand{\TableSep}[2]{\begin{table}[p]\vspace{#1}
\TabCap{#2}\end{table}}

\newcommand{\FigCap}[1]{\footnotesize\par\noindent Fig.\  %
  \refstepcounter{figure}\thefigure. #1\par}

\newcommand{\TableFont}{\footnotesize}
\newcommand{\TableFontIt}{\ttit}
\newcommand{\SetTableFont}[1]{\renewcommand{\TableFont}{#1}}

\newcommand{\MakeTable}[4]{\begin{table}[htb]\TabCap{#2}{#3}
  \begin{center} \TableFont \begin{tabular}{#1} #4 
  \end{tabular}\end{center}\end{table}}

\newcommand{\MakeTableSep}[4]{\begin{table}[p]\TabCap{#2}{#3}
  \begin{center} \TableFont \begin{tabular}{#1} #4 
  \end{tabular}\end{center}\end{table}}

\newenvironment{references}%
{
\footnotesize \frenchspacing
\renewcommand{\thesection}{}
\renewcommand{\in}{{\rm in }}
\renewcommand{\AA}{Astron.\ Astrophys.}
\newcommand{\AAS}{Astron.~Astrophys.~Suppl.~Ser.}
\newcommand{\ApJ}{Astrophys.\ J.}
\newcommand{\ApJS}{Astrophys.\ J.~Suppl.~Ser.}
\newcommand{\ApJL}{Astrophys.\ J.~Letters}
\newcommand{\AJ}{Astron.\ J.}
\newcommand{\IBVS}{IBVS}
\newcommand{\PASP}{P.A.S.P.}
\newcommand{\Acta}{Acta Astron.}
\newcommand{\MNRAS}{MNRAS}
\renewcommand{\and}{{\rm and }}
\section{{\rm REFERENCES}}
\sloppy \hyphenpenalty10000
\begin{list}{}{\leftmargin1cm\listparindent-1cm
\itemindent\listparindent\parsep0pt\itemsep0pt}}%
{\end{list}\vspace{2mm}}

\def\TYLDA{~}
\newlength{\DW}
\settowidth{\DW}{0}
\newcommand{\dw}{\hspace{\DW}}

\newcommand{\refitem}[5]{\item[]{#1} #2%
\def\REFARG{#3}\ifx\REFARG\TYLDA\else, {\it#3}\fi
\def\REFARG{#4}\ifx\REFARG\TYLDA\else, {\bf#4}\fi
\def\REFARG{#5}\ifx\REFARG\TYLDA\else, {#5}\fi.}

\newcommand{\Section}[1]{\section{#1}}
\newcommand{\Subsection}[1]{\subsection{#1}}
\newcommand{\Acknow}[1]{\par\vspace{5mm}{\bf Acknowledgements.} #1}
\pagestyle{myheadings}

\newfont{\bb}{ptmbi8t at 12pt}
\newcommand{\xrule}{\rule{0pt}{2.5ex}}
\newcommand{\xxrule}{\rule[-1.8ex]{0pt}{4.5ex}}
\def\thefootnote{\fnsymbol{footnote}}
\begin{center}
{\Large\bf The Optical Gravitational Lensing Experiment.\\
\vskip3pt
Cepheids in the Galaxy IC1613: No Dependence of the Period--Luminosity
Relation on Metallicity
\footnote{Based on observations obtained
with the 1.3~m Warsaw telescope at the Las Campanas Observatory of the
Carnegie Institution of Washington.}}
\vskip1cm
{\bf A.~~U~d~a~l~s~k~i$^1$,~~\L.~~W~y~r~z~y~k~o~w~s~k~i$^1$,\\
~~G.~~P~i~e~t~r~z~y~\'n~s~k~i$^{2,1}$,~~O.~~S~z~e~w~c~z~y~k$^1$,\\
~~M.~~S~z~y~m~a~{\'n}~s~k~i$^1$,~~M.~~K~u~b~i~a~k$^1$,\\
~ I.~~S~o~s~z~y~\'n~s~k~i$^1$,~~and~~K.~~\.Z~e~b~r~u~\'n$^1$}
\vskip3mm
$^1$Warsaw University Observatory, Al.~Ujazdowskie~4, 00-478~Warszawa, Poland\\
e-mail: (udalski,wyrzykow,pietrzyn,szewczyk,msz,mk,soszynsk,zebrun)@astrouw.edu.pl\\
$^2$ Universidad de Concepci{\'o}n, Departamento de Fisica,
Casilla 160--C, Concepci{\'o}n, Chile
\end{center}

\vspace*{4pt}
\Abstract{
We present results of the search for Cepheids in the galaxy IC1613
carried out as a sub-project of the OGLE-II microlensing survey. 138
Cepheids were found in the $14\zdot\arcm2\times 14\zdot\arcm2$ region 
in the center of the galaxy. We present light curves, {\it VI}
photometry and basic data for all these objects, as well as
color-magnitude diagram of the observed field.

The Period--Luminosity (PL) diagrams for IC1613 fundamental mode
Cepheids for {\it V}, {\it I} and interstellar extinction insensitive
index $W_I$ are constructed. Comparison of PL relations in  metal poor
galaxy IC1613 (${\rm[Fe/H]}\approx-1.0$~dex) with relations in metal
richer Magellanic Clouds allows us to study dependence of Cepheid PL
relations on metallicity in the wide range of metallicities covered by
these three galaxies. The slopes of  PL relations in IC1613 are
identical as in the Magellanic Clouds. The comparison of brightness of
Cepheids with the magnitudes of the tip of the red giant branch stars
and RR Lyr stars in all three objects provides information on the
stability of zero points of PL relations in the {\it I} and {\it
V}-band, respectively. We find that the zero points of PL relations are
constant to better than $\pm0.03$~mag over the entire range of covered
metallicities. Thus, the  most important conclusion of the paper is that
the Cepheid PL relations do not depend on metallicity.

Additionally we determine the mean distance to IC1613, based on the
common distance scale resulting from four major stellar distance
indicators: Cepheids, RR Lyr, TRGB and red clump stars. The distance
modulus to IC1613 is equal to $(m-M)_{\rm IC1613}=24.20$~mag with the
standard deviation of $\pm0.02$~mag from six measurements and systematic
uncertainty of calibrations of $\pm0.07$~mag. 

}{}

\Section{Introduction}

IC1613 is an intrinsically faint irregular galaxy from the Local Group.
It has drawn attention of astronomers for many decades, among others,
because of its favorable location on the sky, far from the Galactic
plane and therefore small contamination by the interstellar reddening.
First search for variable stars in IC1613 was conducted by Baade in
1930s, but its results were published almost four decades later by
Sandage (1971). 37 Classical Cepheids were found based on this
photographic material allowing first determination of the distance to
IC1613 using the Period--Luminosity (PL) relation for Cepheids.
Surprisingly, it turned out that the slope of the PL relation in IC1613
was significantly smaller than in other galaxies possessing Cepheids.
This led to long lasting  discussions on the possible reason of this
discrepancy which, if real, would question usefulness of the PL relation
for distance determinations. The problem was, at least partially, solved
by Freedman (1988a). Much more precise CCD photometry of Cepheids from
IC1613 (however, a small sample of only 11 objects) seemed to contradict
possibility of different slope of the PL relation in IC1613.
 
More recently astronomers realized that IC1613 can be a very good
laboratory for testing properties of standard candles. Beside of
Cepheids, the galaxy contains also the tip of the red giant branch
stars (Freedman 1988b) and RR~Lyr stars (Saha \etal 1992) reachable from
ground based observatories. However, for better quality observations of
RR~Lyr and another standard candle red clump stars  HST telescope must
be used (Dolphin \etal 2001).

It was also realized that IC1613 is a low metallicity object  (Freedman
1988b) with metallicity of about 0.3~dex lower than that of the well
studied SMC. Therefore, IC1613 may be a crucial object for testing
properties of standard candles in low metallicity environment and for
calibrating, if present, population effects on their brightness. In
particular, precise photometry of Cepheids in IC1613 may shed a light on
the possible dependence of the PL relations on metallicity. There is no
general agreement in this matter: theoretical modeling  leads to
contradictory results (Saio and Gautschy 1998, Alibert \etal 1999, Bono
\etal 1999, Sandage, Bell and Tripicco 1999, Caputo \etal 2000). On the
other hand the empirical attempts to solve this problem were also
non-conclusive and with high degree of uncertainty (Sasselov \etal 1997,
Kochanek 1997, Kennicutt \etal 1998). The problem is extremely
important, as Cepheid PL relations play the basic role in establishing
the extragalactic distance scale.

While ground observations of the red clump and RR~Lyr stars require the
largest telescopes, it is, however, somewhat surprising that practically
no wide extensive search for Cepheids in IC1613 has been undertaken
since the Baade's survey in 1930s. Cepheids in IC1613 are reachable from
the ground by the medium size telescopes.  The only exception here was
the survey reported by  Antonello \etal (1999, 2000) and Mantegazza \etal
(2001) who discovered about 100 new Cepheids in four fields in IC1613.
However, their survey was conducted   in white light making their data
of little use for studying the Cepheid PL relations and other problems
requiring well calibrated photometry.

In this paper we present the results of a sub-project of the OGLE-II
microlensing survey (Udalski, Kubiak and Szyma{\'n}ski 1997), which was
undertaken to fill this gap. We report detection of 138 Cepheids in
IC1613 and provide their standard {\it VI}-band CCD photometry. We show
the PL relations for Cepheids in IC1613 and analyze them by comparison
with PL relations in the Magellanic Clouds, and comparison of brightness
of Cepheids with brightness of other stellar standard candles. The main
conclusion of the paper is that the PL relations for the fundamental
mode classical Cepheids are universal and do not depend on metallicity.

Additionally we determine the distance to IC1613 based on photometry of
standard candles presented in this paper or taken from the literature
(Dolphin \etal 2001) and calibrations of the common distance scale of
major stellar distance indicators provided by Udalski (2000b).

All photometric data presented in this paper are available from the OGLE
Internet archive.

\vspace*{12pt}
\Section{Observational Data}

Observations of IC1613 galaxy  presented in this paper were  carried out
as a sub-project of the second phase of the OGLE microlensing search.
The  1.3-m Warsaw telescope at the Las Campanas Observatory, Chile,
(operated by the Carnegie Institution of Washington) was used. The
telescope was equipped  with the "first generation" camera with a SITe
${2048\times2048}$ CCD detector  working in still-frame mode. The pixel
size was 24~$\mu$m giving the 0.417  arcsec/pixel scale. Observations
were performed in the ``medium'' reading mode of  the CCD detector with
the gain 7.1~e$^-$/ADU and readout noise of about  6.3~e$^-$. Details of
the instrumentation setup can be found in Udalski, Kubiak and
Szyma{\'n}ski (1997). 

The photometric data were collected on 51 nights between August 30 and
November 26, 2000. One field centered on the galaxy center
(RA=1\uph04\upm50\ups, DEC=$2\arcd08\arcm00\arcs$, 2000.0) covering
$14\zdot\arcm2\times 14\zdot\arcm2$ on the sky was observed.  One set of
{\it V} and {\it I}-band observations was obtained on each night. The
exposure time was equal to 900 seconds for both {\it V} and {\it
I}-band. Unfortunately, after October~6, 2000 it was not possible to
obtain so long exposures because of the autoguider failure. Since then,
each {\it V} and {\it I}-band observation consisted of eight 120 seconds
unguided exposures which were then stacked and summed, giving the
effective exposure time of 960 seconds. 

Observations were usually made at good atmospheric and seeing
conditions. However, a few worst quality images obtained at bad seeing
and/or with passing clouds were removed from our final dataset. The
median seeing of the analyzed images is 1\zdot\arcs2 and 1\zdot\arcs1
for the {\it V} and {\it I}-band, respectively.

On seven nights when observations IC1613 were obtained, several standard
stars from the Landolt (1992) list were also observed during the regular
OGLE-II survey. They were used to transform photometry of IC1613 to the
standard system.

\Section{Data Reduction}

Collected data were de-biased and flat-fielded in the real time at the
telescope using the standard OGLE-II data pipeline (Udalski, Kubiak and
Szyma{\'n}ski 1997). Before proceeding to further reductions, the
observations consisting of eight short exposures were processed. First,
seven images of the set were resampled to the first (reference) image 
pixel coordinate system using procedures from the DIA package of
Wo\'zniak (2000). Then, the intensity scale of these images was fitted
to the scale of the reference frame and all stack of images was coadded
using standard {\sc IRAF}\footnote{IRAF is distributed by National
Optical Observatories, which is operated by the Association of
Universities for Reaserch in Astronomy, Inc., under cooperative
agreement with National Science Foundation.} package procedures. This
procedure worked very effectively, no significant seeing or photometric
quality degradation was noted as compared to the regular long exposure
images.

The photometry of stars was derived using the {\sc DoPhot} photometry
package (Schechter, Mateo and Saha 1993) run on  $512\times512$ pixel
subframes to account for Point Spread Function (PSF) variations over the
frame. Two very good seeing frames -- one for {\it V} and the other for
{\it I}-band, were used as the template images whose photometry defined
the instrumental {\it VI} system. Photometry of each of the remaining
images was compared then to the template one and shifted appropriately
to the instrumental system.

In the next stage, the photometric {\it V} and {\it I}-band databases
containing photometry obtained from all images were created. They were
then calibrated to the standard system by adding corrections resulting
from observations of standard stars and  the aperture correction to the
instrumental photometry averaged from several determinations. Finally,
small corrections resulting from imperfect flat-fielding procedure of
the standard OGLE-II data pipeline (Udalski 2000b) and mapped from
observations of  stars of known brightness were applied. The total error
of calibration should not exceed 0.02 mag for both {\it V} and {\it
I}-band.

\Section{Cepheids in IC1613}

The IC1613 photometric data in the {\it VI} databases were searched for
periodic variable objects. All stars with standard deviation of
brightness larger than typical for constant brightness objects were
selected as potential variable candidates. Their light curves were then
searched for periodicity using the AoV algorithm (Schwarzenberg-Czerny
1989). Because the number of collected epochs was similar in the {\it V}
and {\it I}-bands, the search for periodic objects was performed
independently for both these bands.

Cepheids in IC1613 were selected from the list of periodic variable
objects, by visual inspection of all phased light curves. The main
criterion of Cepheid classification was the very characteristic light
curve shape and the period in the range of about 1 to 50 days. The
latter limit resulted from the total duration of our observations.

\renewcommand{\TableFont}{\tiny}
\renewcommand{\arraystretch}{0.55}
\setcounter{table}{0}
\MakeTableSep{r@{\hspace{5pt}}
           c@{\hspace{3pt}}
           c@{\hspace{3pt}}
           r
           c@{\hspace{3pt}}
           c@{\hspace{3pt}}
           c@{\hspace{3pt}}
           c@{\hspace{3pt}}
           c@{\hspace{3pt}}
}{13cm}
{Cepheids in IC1613}
{
\hline
\noalign{\vskip2pt}
\multicolumn{1}{c}{Star} & RA & DEC &\multicolumn{1}{c}{$P$} & 
$I$ & $V$ & $V-I$ & $W_I$ & \multicolumn{1}{c}{Remarks}\\
\noalign{\vskip1pt}
\multicolumn{1}{c}{number} & (J2000) & (J2000) & 
\multicolumn{1}{c}{[days]} & [mag] & 
[mag] & [mag] & [mag] &  \\
\noalign{\vskip2pt}
\hline
\noalign{\vskip3pt}
 11446 & 1\uph04\upm59\zdot\ups84 & 2\arcd05\arcm28\zdot\arcs0 & 41.630 & 17.855 & 18.801 &  0.945 & 16.390 & V20 \\
 10421 & 1\uph04\upm58\zdot\ups12 & 2\arcd02\arcm32\zdot\arcs8 & 29.310 & 20.389 & 21.099 &  0.710 & 19.289 & V47, PII \\
  1987 & 1\uph04\upm31\zdot\ups81 & 2\arcd10\arcm06\zdot\arcs7 & 25.862 & 18.554 & 19.393 &  0.838 & 17.254 & V11 \\
   736 & 1\uph04\upm32\zdot\ups23 & 2\arcd05\arcm01\zdot\arcs6 & 23.450 & 18.446 & 19.258 &  0.811 & 17.189 & V2 \\
  7647 & 1\uph04\upm37\zdot\ups80 & 2\arcd09\arcm08\zdot\arcs1 & 16.540 & 18.515 & 18.994 &  0.478 & 17.774 & \\
 13738 & 1\uph05\upm02\zdot\ups91 & 2\arcd10\arcm34\zdot\arcs8 & 16.370 & 19.113 & 19.988 &  0.874 & 17.758 & V18 \\
 13682 & 1\uph05\upm02\zdot\ups15 & 2\arcd10\arcm24\zdot\arcs3 & 14.330 & 17.617 & 18.851 &  1.233 & 15.705 & V39 \\
 17473 & 1\uph05\upm05\zdot\ups86 & 2\arcd07\arcm34\zdot\arcs5 & 13.120 & 21.498 & 21.887 &  0.388 & 20.896 & PII \\
  4861 & 1\uph04\upm44\zdot\ups34 & 2\arcd05\arcm29\zdot\arcs0 & 12.410 & 19.451 & 20.284 &  0.832 & 18.163 & V37 \\
  7664 & 1\uph04\upm41\zdot\ups52 & 2\arcd08\arcm23\zdot\arcs9 & 10.450 & 19.425 & 20.064 &  0.638 & 18.437 & V16 \\
   926 & 1\uph04\upm33\zdot\ups69 & 2\arcd07\arcm45\zdot\arcs3 &  9.402 & 19.637 & 20.296 &  0.658 & 18.617 & V6 \\
   879 & 1\uph04\upm34\zdot\ups54 & 2\arcd06\arcm42\zdot\arcs1 &  9.193 & 19.601 & 20.241 &  0.639 & 18.611 & V25 \\
 11589 & 1\uph04\upm51\zdot\ups61 & 2\arcd05\arcm33\zdot\arcs2 &  8.409 & 19.997 & 20.729 &  0.731 & 18.865 & V34 \\
 13808 & 1\uph04\upm59\zdot\ups84 & 2\arcd08\arcm42\zdot\arcs8 &  7.557 & 20.175 & 20.945 &  0.769 & 18.984 & \\
 18919 & 1\uph05\upm06\zdot\ups70 & 2\arcd12\arcm52\zdot\arcs4 &  7.551 & 19.889 &   --   &   --   &   --   & V49 \\
 13759 & 1\uph04\upm52\zdot\ups61 & 2\arcd08\arcm04\zdot\arcs5 &  7.333 & 19.987 &   --   &   --   &   --   & V7 \\
 18905 & 1\uph05\upm06\zdot\ups41 & 2\arcd12\arcm33\zdot\arcs6 &  6.766 & 20.226 & 21.018 &  0.792 & 18.998 & \\
 13943 & 1\uph04\upm51\zdot\ups77 & 2\arcd10\arcm54\zdot\arcs7 &  6.751 & 19.960 & 20.502 &  0.541 & 19.122 & V24 \\
 13709 & 1\uph04\upm57\zdot\ups03 & 2\arcd08\arcm40\zdot\arcs7 &  6.748 & 19.464 & 20.556 &  1.091 & 17.773 & \\
  3732 & 1\uph04\upm40\zdot\ups31 & 2\arcd01\arcm24\zdot\arcs5 &  6.669 & 20.064 & 20.680 &  0.616 & 19.110 & V27 \\
  5037 & 1\uph04\upm49\zdot\ups24 & 2\arcd07\arcm19\zdot\arcs9 &  6.310 & 20.294 &   --   &   --   &   --   & \\
 17454 & 1\uph05\upm16\zdot\ups39 & 2\arcd07\arcm21\zdot\arcs8 &  6.111 & 20.778 &   --   &   --   &   --   & \\
 11604 & 1\uph04\upm56\zdot\ups72 & 2\arcd05\arcm48\zdot\arcs0 &  5.885 & 20.490 & 21.271 &  0.780 & 19.281 & \\
  3722 & 1\uph04\upm43\zdot\ups93 & 2\arcd01\arcm04\zdot\arcs4 &  5.818 & 20.294 & 20.994 &  0.700 & 19.209 & V26 \\
 17951 & 1\uph05\upm04\zdot\ups79 & 2\arcd08\arcm50\zdot\arcs3 &  5.738 & 20.013 &   --   &   --   &   --   & \\
 13911 & 1\uph04\upm51\zdot\ups70 & 2\arcd10\arcm10\zdot\arcs2 &  5.717 & 20.169 & 20.685 &  0.516 & 19.369 & V17 \\
 13780 & 1\uph04\upm56\zdot\ups35 & 2\arcd08\arcm21\zdot\arcs3 &  5.580 & 20.367 & 20.991 &  0.623 & 19.401 & V9 \\
  4875 & 1\uph04\upm49\zdot\ups08 & 2\arcd05\arcm36\zdot\arcs8 &  5.138 & 20.245 & 20.918 &  0.672 & 19.204 & V14 \\
 11831 & 1\uph04\upm57\zdot\ups54 & 2\arcd04\arcm44\zdot\arcs4 &  5.028 & 20.757 & 21.472 &  0.714 & 19.650 & \\
 15696 & 1\uph04\upm51\zdot\ups03 & 2\arcd14\arcm30\zdot\arcs3 &  5.012 & 20.565 & 21.245 &  0.679 & 19.513 & \\
 15670 & 1\uph04\upm53\zdot\ups39 & 2\arcd13\arcm30\zdot\arcs3 &  4.849 & 20.386 & 20.956 &  0.569 & 19.504 & V13 \\
  5574 & 1\uph04\upm50\zdot\ups00 & 2\arcd06\arcm01\zdot\arcs4 &  4.829 & 20.723 & 21.512 &  0.789 & 19.500 & \\
 17805 & 1\uph05\upm12\zdot\ups48 & 2\arcd07\arcm13\zdot\arcs6 &  4.739 & 20.922 & 21.683 &  0.760 & 19.744 & \\
  8146 & 1\uph04\upm37\zdot\ups52 & 2\arcd08\arcm51\zdot\arcs2 &  4.568 & 20.870 & 21.558 &  0.687 & 19.805 & \\
 14287 & 1\uph05\upm01\zdot\ups15 & 2\arcd09\arcm11\zdot\arcs5 &  4.365 & 20.893 & 21.655 &  0.761 & 19.714 & \\
 18891 & 1\uph05\upm13\zdot\ups08 & 2\arcd12\arcm12\zdot\arcs5 &  4.287 & 20.679 & 21.300 &  0.620 & 19.717 & V12 \\
 12415 & 1\uph05\upm00\zdot\ups12 & 2\arcd06\arcm59\zdot\arcs8 &  4.264 & 20.802 & 21.428 &  0.625 & 19.833 & \\
  7919 & 1\uph04\upm49\zdot\ups10 & 2\arcd08\arcm02\zdot\arcs5 &  4.264 & 20.729 & 21.403 &  0.673 & 19.686 & V30 \\
  5857 & 1\uph04\upm47\zdot\ups97 & 2\arcd06\arcm48\zdot\arcs9 &  4.218 & 20.520 & 21.201 &  0.680 & 19.467 & V15 \\
 12109 & 1\uph05\upm01\zdot\ups31 & 2\arcd05\arcm52\zdot\arcs9 &  4.132 & 20.757 & 21.359 &  0.602 & 19.823 & \\
 13784 & 1\uph04\upm59\zdot\ups96 & 2\arcd08\arcm24\zdot\arcs7 &  4.045 & 20.451 & 21.046 &  0.595 & 19.528 & V10 \\
  5309 & 1\uph04\upm43\zdot\ups24 & 2\arcd05\arcm19\zdot\arcs9 &  4.029 & 20.618 & 21.414 &  0.796 & 19.384 & \\
  4840 & 1\uph04\upm43\zdot\ups20 & 2\arcd05\arcm17\zdot\arcs5 &  4.013 &   --   & 21.310 &   --   &   --   & V61 \\
  1132 & 1\uph04\upm23\zdot\ups62 & 2\arcd06\arcm11\zdot\arcs7 &  4.009 & 20.834 & 21.527 &  0.693 & 19.759 & \\
  5256 & 1\uph04\upm48\zdot\ups33 & 2\arcd05\arcm05\zdot\arcs5 &  4.003 & 20.875 &   --   &   --   &   --   & V3 \\
 11743 & 1\uph04\upm51\zdot\ups44 & 2\arcd07\arcm32\zdot\arcs3 &  3.893 & 20.007 & 20.830 &  0.822 & 18.733 & V53 \\
  6084 & 1\uph04\upm46\zdot\ups55 & 2\arcd07\arcm28\zdot\arcs1 &  3.872 & 20.614 & 21.471 &  0.856 & 19.288 & \\
  8127 & 1\uph04\upm40\zdot\ups38 & 2\arcd08\arcm48\zdot\arcs1 &  3.856 & 20.863 & 21.606 &  0.742 & 19.712 & \\
 18222 & 1\uph05\upm06\zdot\ups02 & 2\arcd09\arcm50\zdot\arcs2 &  3.803 & 21.033 & 21.634 &  0.601 & 20.101 & \\
 14356 & 1\uph05\upm02\zdot\ups44 & 2\arcd09\arcm30\zdot\arcs1 &  3.666 & 20.819 & 21.439 &  0.620 & 19.859 & V54 \\
   961 & 1\uph04\upm27\zdot\ups58 & 2\arcd04\arcm38\zdot\arcs2 &  3.638 & 21.054 & 21.812 &  0.757 & 19.881 & \\
  9928 & 1\uph04\upm47\zdot\ups19 & 2\arcd14\arcm21\zdot\arcs6 &  3.608 & 20.572 &   --   &   --   &   --   & V57 \\
  5614 & 1\uph04\upm45\zdot\ups33 & 2\arcd06\arcm08\zdot\arcs6 &  3.434 & 21.159 & 21.904 &  0.744 & 20.005 & \\
  5611 & 1\uph04\upm43\zdot\ups76 & 2\arcd06\arcm08\zdot\arcs7 &  3.255 & 20.960 & 21.629 &  0.668 & 19.925 & \\
  5750 & 1\uph04\upm44\zdot\ups52 & 2\arcd06\arcm30\zdot\arcs7 &  3.247 & 21.122 & 21.796 &  0.673 & 20.079 & \\
  5281 & 1\uph04\upm38\zdot\ups39 & 2\arcd05\arcm11\zdot\arcs9 &  3.144 & 20.640 & 21.249 &  0.609 & 19.696 & \\
  2240 & 1\uph04\upm34\zdot\ups84 & 2\arcd09\arcm07\zdot\arcs9 &  3.074 & 21.096 & 21.608 &  0.512 & 20.302 & V35 \\
   342 & 1\uph04\upm25\zdot\ups66 & 2\arcd03\arcm53\zdot\arcs8 &  3.065 & 21.089 & 21.711 &  0.621 & 20.126 & V62 \\
  6273 & 1\uph04\upm50\zdot\ups43 & 2\arcd07\arcm57\zdot\arcs7 &  3.019 & 21.109 & 21.797 &  0.687 & 20.044 & \\
  4016 & 1\uph04\upm40\zdot\ups08 & 2\arcd03\arcm17\zdot\arcs5 &  2.957 & 21.450 & 22.150 &  0.700 & 20.365 & \\
 18349 & 1\uph05\upm05\zdot\ups34 & 2\arcd11\arcm06\zdot\arcs9 &  2.869 & 21.142 & 21.698 &  0.556 & 20.280 & V29 \\
 19024 & 1\uph05\upm06\zdot\ups17 & 2\arcd11\arcm52\zdot\arcs3 &  2.843 &   --   & 21.858 &   --   &   --   & \\
 16011 & 1\uph04\upm55\zdot\ups30 & 2\arcd14\arcm49\zdot\arcs9 &  2.795 &   --   & 22.100 &   --   &   --   & \\
 12233 & 1\uph04\upm50\zdot\ups83 & 2\arcd06\arcm20\zdot\arcs8 &  2.794 & 21.175 & 21.830 &  0.654 & 20.160 & \\
 12068 & 1\uph04\upm53\zdot\ups92 & 2\arcd05\arcm41\zdot\arcs1 &  2.781 &   --   & 21.872 &   --   &   --   & \\
  7232 & 1\uph04\upm37\zdot\ups16 & 2\arcd07\arcm01\zdot\arcs7 &  2.774 & 21.306 & 21.975 &  0.668 & 20.271 & \\
  2760 & 1\uph04\upm33\zdot\ups89 & 2\arcd09\arcm18\zdot\arcs5 &  2.712 & 21.297 & 21.870 &  0.572 & 20.410 & \\
  1471 & 1\uph04\upm23\zdot\ups91 & 2\arcd05\arcm17\zdot\arcs6 &  2.707 & 21.320 & 21.958 &  0.637 & 20.333 & \\
  7010 & 1\uph04\upm48\zdot\ups63 & 2\arcd06\arcm30\zdot\arcs2 &  2.673 & 21.273 & 21.889 &  0.616 & 20.319 & \\
}

\setcounter{table}{0}
\MakeTableSep{r@{\hspace{5pt}}
           c@{\hspace{3pt}}
           c@{\hspace{3pt}}
           r
           c@{\hspace{3pt}}
           c@{\hspace{3pt}}
           c@{\hspace{3pt}}
           c@{\hspace{3pt}}
           c@{\hspace{3pt}}
}{13cm}
{(concluded)}
{
\hline
\noalign{\vskip2pt}
\multicolumn{1}{c}{Star} & RA & DEC &\multicolumn{1}{c}{$P$} & 
$I$ & $V$ & $V-I$ & $W_I$ & \multicolumn{1}{c}{Remarks}\\
\noalign{\vskip1pt}
\multicolumn{1}{c}{number} & (J2000) & (J2000) & 
\multicolumn{1}{c}{[days]} & [mag] & 
[mag] & [mag] & [mag] &  \\
\noalign{\vskip2pt}
\hline
\noalign{\vskip3pt}
  1028 & 1\uph04\upm28\zdot\ups26 & 2\arcd05\arcm16\zdot\arcs2 &  2.664 & 21.480 & 22.146 &  0.665 & 20.449 & \\
 10804 & 1\uph04\upm51\zdot\ups05 & 2\arcd04\arcm08\zdot\arcs0 &  2.662 & 21.312 & 21.812 &  0.499 & 20.540 & V48 \\
  4080 & 1\uph04\upm41\zdot\ups01 & 2\arcd03\arcm49\zdot\arcs5 &  2.632 & 20.934 & 21.727 &  0.793 & 19.705 & V51 \\
 12526 & 1\uph05\upm02\zdot\ups97 & 2\arcd07\arcm23\zdot\arcs7 &  2.631 &   --   & 22.231 &   --   &   --   & \\
 10263 & 1\uph04\upm36\zdot\ups77 & 2\arcd14\arcm11\zdot\arcs2 &  2.566 & 21.320 & 21.967 &  0.647 & 20.317 & V46 \\
  6603 & 1\uph04\upm48\zdot\ups56 & 2\arcd05\arcm30\zdot\arcs2 &  2.564 & 21.584 & 22.356 &  0.771 & 20.390 & \\
 16245 & 1\uph05\upm00\zdot\ups75 & 2\arcd13\arcm13\zdot\arcs9 &  2.561 & 21.396 & 21.950 &  0.554 & 20.537 & \\
  4160 & 1\uph04\upm42\zdot\ups16 & 2\arcd04\arcm15\zdot\arcs0 &  2.542 & 21.484 & 22.136 &  0.651 & 20.474 & \\
 16301 & 1\uph04\upm55\zdot\ups64 & 2\arcd13\arcm42\zdot\arcs6 &  2.533 & 21.492 & 22.289 &  0.797 & 20.257 & \\
 11613 & 1\uph04\upm55\zdot\ups51 & 2\arcd05\arcm52\zdot\arcs0 &  2.476 & 20.604 & 21.018 &  0.414 & 19.962 & \\
 15308 & 1\uph04\upm51\zdot\ups20 & 2\arcd10\arcm08\zdot\arcs3 &  2.460 &   --   & 22.221 &   --   &   --   & \\
 13184 & 1\uph04\upm50\zdot\ups69 & 2\arcd06\arcm22\zdot\arcs7 &  2.455 & 21.381 & 22.006 &  0.624 & 20.413 & \\
 15356 & 1\uph04\upm53\zdot\ups34 & 2\arcd10\arcm21\zdot\arcs9 &  2.447 &   --   & 22.107 &   --   &   --   & V36 \\
 14070 & 1\uph04\upm58\zdot\ups56 & 2\arcd08\arcm23\zdot\arcs2 &  2.396 &   --   & 21.537 &   --   &   --   & \\
  1440 & 1\uph04\upm36\zdot\ups05 & 2\arcd05\arcm05\zdot\arcs6 &  2.339 &   --   & 22.155 &   --   &   --   & \\
  7322 & 1\uph04\upm37\zdot\ups66 & 2\arcd07\arcm13\zdot\arcs7 &  2.338 & 21.347 & 21.944 &  0.597 & 20.421 & \\
  4154 & 1\uph04\upm41\zdot\ups21 & 2\arcd04\arcm12\zdot\arcs6 &  2.267 & 20.932 & 21.508 &  0.575 & 20.041 & V59 \\
 12659 & 1\uph04\upm50\zdot\ups93 & 2\arcd07\arcm54\zdot\arcs9 &  2.265 & 21.463 & 22.031 &  0.567 & 20.584 & \\
   202 & 1\uph04\upm34\zdot\ups73 & 2\arcd02\arcm09\zdot\arcs6 &  2.264 & 21.493 & 22.050 &  0.557 & 20.630 & V28 \\
  6128 & 1\uph04\upm41\zdot\ups32 & 2\arcd07\arcm34\zdot\arcs4 &  2.260 &   --   & 21.608 &   --   &   --   & \\
 14785 & 1\uph04\upm56\zdot\ups12 & 2\arcd08\arcm17\zdot\arcs4 &  2.236 & 21.567 & 22.012 &  0.444 & 20.879 & \\
  1092 & 1\uph04\upm29\zdot\ups58 & 2\arcd05\arcm46\zdot\arcs9 &  2.233 & 21.398 & 21.793 &  0.394 & 20.788 & \\
  4602 & 1\uph04\upm39\zdot\ups25 & 2\arcd04\arcm08\zdot\arcs9 &  2.231 & 21.457 & 22.077 &  0.620 & 20.497 & \\
 14710 & 1\uph05\upm03\zdot\ups25 & 2\arcd08\arcm04\zdot\arcs4 &  2.228 &   --   & 22.406 &   --   &   --   & \\
 15476 & 1\uph04\upm53\zdot\ups35 & 2\arcd11\arcm02\zdot\arcs8 &  2.180 & 21.568 & 22.197 &  0.628 & 20.594 & \\
  6168 & 1\uph04\upm46\zdot\ups59 & 2\arcd07\arcm40\zdot\arcs2 &  2.163 &   --   & 21.544 &   --   &   --   & \\
  2117 & 1\uph04\upm30\zdot\ups25 & 2\arcd08\arcm06\zdot\arcs5 &  2.131 & 21.300 & 21.857 &  0.557 & 20.437 & \\
  5209 & 1\uph04\upm43\zdot\ups02 & 2\arcd04\arcm53\zdot\arcs4 &  2.094 &   --   & 21.892 &   --   &   --   & \\
  8782 & 1\uph04\upm36\zdot\ups87 & 2\arcd08\arcm22\zdot\arcs6 &  2.091 & 21.480 & 22.044 &  0.563 & 20.607 & \\
  5996 & 1\uph04\upm44\zdot\ups84 & 2\arcd07\arcm14\zdot\arcs1 &  2.069 &   --   & 21.666 &   --   &   --   & V60 \\
  2389 & 1\uph04\upm28\zdot\ups68 & 2\arcd10\arcm24\zdot\arcs5 &  2.029 &   --   & 21.571 &   --   &   --   & \\
 12044 & 1\uph04\upm52\zdot\ups52 & 2\arcd05\arcm35\zdot\arcs3 &  2.027 & 21.272 & 21.786 &  0.514 & 20.475 & \\
 12747 & 1\uph05\upm04\zdot\ups26 & 2\arcd04\arcm47\zdot\arcs8 &  1.976 &   --   & 22.087 &   --   &   --   & \\
  2342 & 1\uph04\upm33\zdot\ups07 & 2\arcd09\arcm56\zdot\arcs9 &  1.972 & 21.203 & 21.751 &  0.548 & 20.353 & \\
 17400 & 1\uph05\upm13\zdot\ups00 & 2\arcd06\arcm47\zdot\arcs8 &  1.969 &   --   & 22.087 &   --   &   --   & \\
 12929 & 1\uph04\upm57\zdot\ups71 & 2\arcd05\arcm30\zdot\arcs7 &  1.941 &   --   & 22.553 &   --   &   --   & \\
 14790 & 1\uph04\upm57\zdot\ups46 & 2\arcd08\arcm18\zdot\arcs2 &  1.913 &   --   & 22.617 &   --   &   --   & \\
  5674 & 1\uph04\upm48\zdot\ups52 & 2\arcd06\arcm18\zdot\arcs1 &  1.888 & 20.829 & 21.448 &  0.619 & 19.870 & \\
  4475 & 1\uph04\upm36\zdot\ups79 & 2\arcd03\arcm32\zdot\arcs1 &  1.818 & 21.793 & 22.361 &  0.567 & 20.914 & \\
 15230 & 1\uph04\upm54\zdot\ups70 & 2\arcd09\arcm47\zdot\arcs0 &  1.765 &   --   & 22.187 &   --   &   --   & \\
  2124 & 1\uph04\upm22\zdot\ups99 & 2\arcd08\arcm14\zdot\arcs0 &  1.697 & 21.585 & 22.177 &  0.591 & 20.669 & \\
 13481 & 1\uph04\upm57\zdot\ups68 & 2\arcd07\arcm25\zdot\arcs6 &  1.678 &   --   & 22.595 &   --   &   --   & \\
  2818 & 1\uph04\upm26\zdot\ups81 & 2\arcd09\arcm38\zdot\arcs8 &  1.661 &   --   & 22.642 &   --   &   --   & \\
 12401 & 1\uph04\upm54\zdot\ups18 & 2\arcd06\arcm56\zdot\arcs1 &  1.657 & 21.450 & 21.979 &  0.529 & 20.631 & \\
  7018 & 1\uph04\upm47\zdot\ups72 & 2\arcd06\arcm31\zdot\arcs7 &  1.645 &   --   & 22.358 &   --   &   --   & \\
 17545 & 1\uph05\upm10\zdot\ups87 & 2\arcd04\arcm39\zdot\arcs4 &  1.623 &   --   & 22.480 &   --   &   --   & \\
 13192 & 1\uph05\upm00\zdot\ups15 & 2\arcd06\arcm23\zdot\arcs7 &  1.620 &   --   & 21.222 &   --   &   --   & \\
  3897 & 1\uph04\upm47\zdot\ups24 & 2\arcd01\arcm53\zdot\arcs9 &  1.618 &   --   & 22.304 &   --   &   --   & \\
  4195 & 1\uph04\upm41\zdot\ups38 & 2\arcd04\arcm26\zdot\arcs8 &  1.603 &   --   & 22.049 &   --   &   --   & \\
 11130 & 1\uph05\upm04\zdot\ups49 & 2\arcd02\arcm57\zdot\arcs5 &  1.599 &   --   & 22.218 &   --   &   --   & \\
  5197 & 1\uph04\upm50\zdot\ups43 & 2\arcd04\arcm51\zdot\arcs6 &  1.588 &   --   & 21.914 &   --   &   --   & \\
 19479 & 1\uph05\upm13\zdot\ups62 & 2\arcd12\arcm04\zdot\arcs5 &  1.565 &   --   & 22.029 &   --   &   --   & \\
  2197 & 1\uph04\upm24\zdot\ups41 & 2\arcd08\arcm46\zdot\arcs4 &  1.459 &   --   & 22.430 &   --   &   --   & \\
 12070 & 1\uph05\upm01\zdot\ups52 & 2\arcd05\arcm40\zdot\arcs7 &  1.438 &   --   & 21.160 &   --   &   --   & \\
  2751 & 1\uph04\upm32\zdot\ups52 & 2\arcd09\arcm17\zdot\arcs0 &  1.376 &   --   & 21.788 &   --   &   --   & \\
  7028 & 1\uph04\upm42\zdot\ups93 & 2\arcd06\arcm33\zdot\arcs5 &  1.375 &   --   & 22.276 &   --   &   --   & \\
  2771 & 1\uph04\upm34\zdot\ups72 & 2\arcd09\arcm22\zdot\arcs3 &  1.329 &   --   & 22.198 &   --   &   --   & \\
  8173 & 1\uph04\upm39\zdot\ups98 & 2\arcd08\arcm58\zdot\arcs0 &  1.310 &   --   & 21.914 &   --   &   --   & \\
  2909 & 1\uph04\upm28\zdot\ups22 & 2\arcd10\arcm07\zdot\arcs3 &  1.309 &   --   & 22.250 &   --   &   --   & \\
 10862 & 1\uph05\upm01\zdot\ups58 & 2\arcd01\arcm05\zdot\arcs5 &  1.225 &   --   & 21.899 &   --   &   --   & \\
  4173 & 1\uph04\upm42\zdot\ups36 & 2\arcd04\arcm19\zdot\arcs4 &  1.199 &   --   & 22.085 &   --   &   --   & \\
  4555 & 1\uph04\upm41\zdot\ups71 & 2\arcd03\arcm57\zdot\arcs4 &  1.167 &   --   & 22.284 &   --   &   --   & \\
 18698 & 1\uph05\upm15\zdot\ups01 & 2\arcd10\arcm20\zdot\arcs1 &  1.156 &   --   & 22.350 &   --   &   --   & \\
 12167 & 1\uph04\upm55\zdot\ups88 & 2\arcd06\arcm04\zdot\arcs8 &  1.113 &   --   & 21.660 &   --   &   --   & \\
 11221 & 1\uph04\upm50\zdot\ups95 & 2\arcd03\arcm27\zdot\arcs9 &  1.097 &   --   & 22.415 &   --   &   --   & \\
  1383 & 1\uph04\upm31\zdot\ups99 & 2\arcd04\arcm39\zdot\arcs4 &  1.021 &   --   & 22.383 &   --   &   --   & \\
 19350 & 1\uph05\upm07\zdot\ups13 & 2\arcd14\arcm44\zdot\arcs1 &  0.850 &   --   & 22.459 &   --   &   --   & \\
 19304 & 1\uph05\upm12\zdot\ups36 & 2\arcd14\arcm14\zdot\arcs9 &  0.822 &   --   & 22.601 &   --   &   --   & \\
 19725 & 1\uph05\upm07\zdot\ups91 & 2\arcd13\arcm46\zdot\arcs4 &  0.723 &   --   & 22.439 &   --   &   --   & \\
}

\renewcommand{\TableFont}{\footnotesize}
\renewcommand{\arraystretch}{1}

Independent search in the {\it V} and {\it I}-band allowed us to find
more complete sample of Cepheids in IC1613 and exclude some doubtful
cases. Especially, deeper range of the {\it V}-band photometry made it
possible to identify many more fainter Cepheids falling into noise in
the {\it I}-band frames.

Altogether we selected 131 light curves in the {\it V}-band and 91 in
the {\it I}-band of 138 Cepheid candidates. Phased light curves of these
objects are presented in Appendix~A. Table~1 provides the most important
observational data for each candidate. In the first column  the
identification number in the OGLE databases is given followed by the
equatorial coordinates (2000.0), period and intensity mean {\it I}, {\it
V}, $(V-I)$ photometry. In the next column the interstellar extinction
independent index (Wesenheit index), $W_I$, defined as:
$$ W_I=I-1.55\cdot(V-I) \eqno{(1)} $$
is also listed. In some cases remarks are provided in the last column.
In particular, cross-identification with Sandage (1971) notation  is
given. Also Population~II Cepheids (PII)  are marked in that column.

The mean intensity photometry was determined by fitting the Fourier
series of the fifth order to the observed light curves. Formal accuracy
of such determined mean magnitudes is of about 0.005~mag  for the
brightest stars to about 0.04 mag for the faintest ones. Accuracy of the
period is about $1\cdot10^{-3}\cdot P$. 

It is difficult to estimate how complete is the presented sample. Large
amplitudes make the Cepheids easy to detect and the sample is very
likely complete close to 100\%  at the bright end. For fainter objects
the completeness certainly decreases because of larger photometric
noise. Also non-uniform background of the galaxy makes detection more
difficult in some regions. Comparison of objects detected in the {\it V}
and {\it I}-bands suggests that down to $P\approx2.5$~days the sample is
reasonably complete -- single objects were present only on one list.
Objects with shorter periods were detected practically in the {\it
V}-band only, and for obvious reasons their completeness must be much
smaller.

IC1613's location on the sky, as well as its orientation are very
fortunate as far as  the interstellar reddening is considered. It is
commonly accepted that the internal reddening in IC1613 is negligible,
and the Galactic foreground reddening in the IC1613 direction is very
small. According to Schlegel, Finkbeiner and Davis (1998) $E(B-V)$
toward the galaxy is only 0.025~mag. Thus, to convert the presented
photometry to extinction free one,  one has to subtract $A_I=0.05$~mag
and $A_V=0.08$~mag for the {\it I} and {\it V}-band, respectively, from
the data in Table~1.

\begin{figure}[p]
\hglue-6mm
\includegraphics[bb=40 65 550 730,width=13cm]{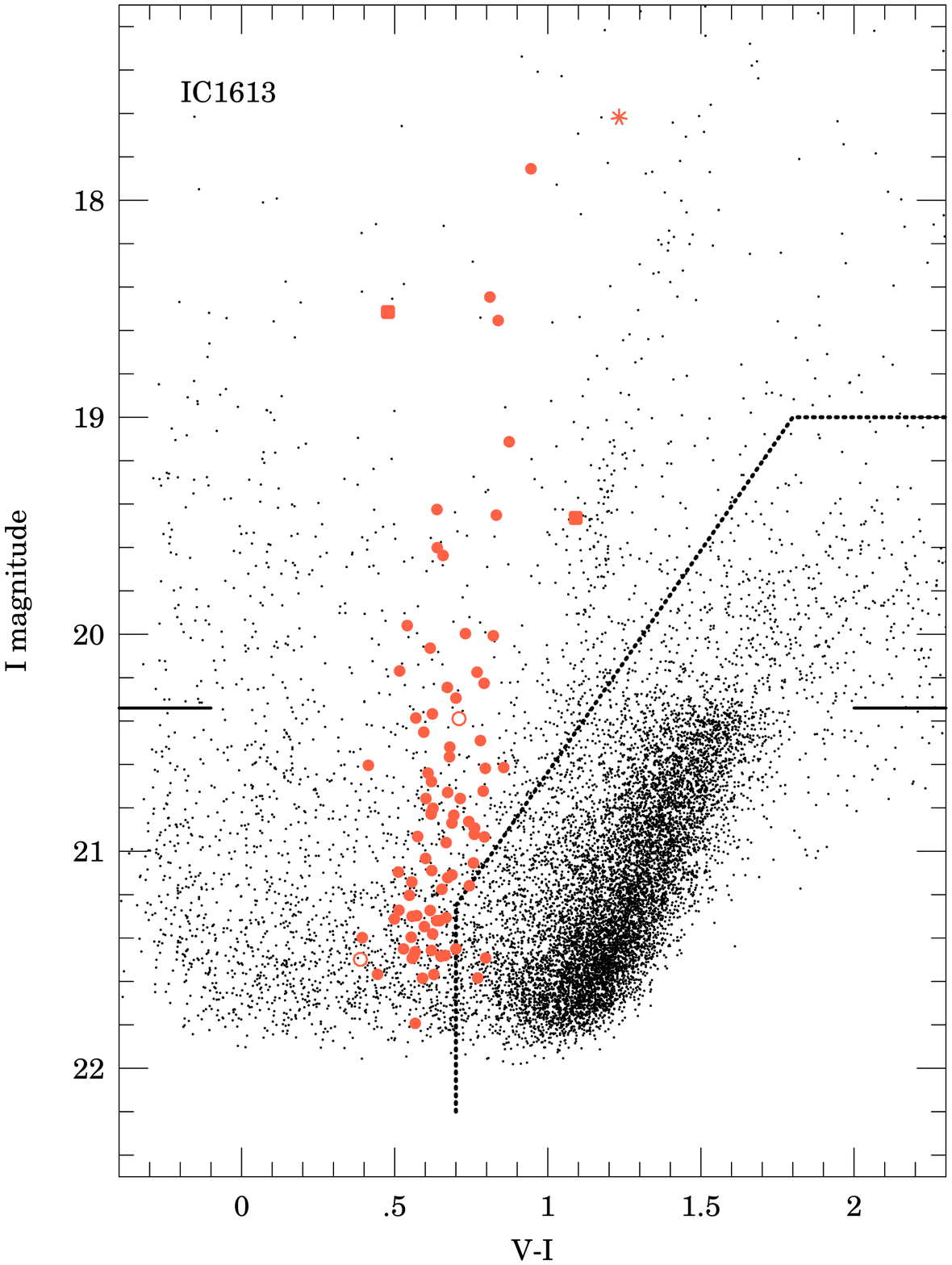}
\vskip3pt
\FigCap{Color-Magnitude diagram of the observed field in IC1613. Large
grey dots indicate positions of classical Cepheids. Gray squares
correspond to the positions of blended Cepheids, 7647 and 13709. Open
circles mark Population~II Cepheids and asterisk an enigmatic variable,
13862. Thick dotted line limits the region used for determination of the
TRGB magnitude which is marked by two thick solid lines on the abscissa
axes at 20.34~mag.}
\end{figure}

Fig.~1 presents the color-magnitude diagram (CMD) for stars from our
photometric databases of IC1613. 80 classical Cepheids from Table~1,
that is those with full {\it VI} photometry, are marked by large grey
dots in this figure. Open grey circles indicate positions of
Population~II Cepheids. Asterisk denotes position of the enigmatic
object, 13682 (V39 according to Sandage 1971) whose light curve
resembles pulsating variable. However, it is very likely not a classical
Cepheid (Sandage 1971, Antonello \etal 1999) what is confirmed here by
its location on the CMD diagram and below on PL diagrams.  The remaining
Cepheids nicely populate the instability strip located left from the
well populated upper part of the red giant branch. Gray squares in
Fig.~1, indicate the classical Cepheids 7647 and 13709. They are located
left or right from the main instability strip and  are certainly
unresolved blends of a Cepheid with another star, what is clearly
visible on the PL diagrams, and from the shape (amplitude) of their {\it
VI} light curves.

\Section{Period--Luminosity Diagrams}

Figs.~2--4 present the PL diagrams for Cepheids in IC1613 plotted using
the data from Table~1, for the {\it V}, {\it I} and $W_I$ index,
respectively. PL diagrams for all bands look very similar. The main
feature is the narrow strip formed by Cepheids pulsating in the
fundamental mode. At the faint end, \ie $\log P < 0.4$ (2.5 days), the
strip becomes  wider and worse defined what is particularly well
noticeable in the {\it V}-band. This is caused by the first overtone
(FO) Cepheids which become much more numerous in this period range and
are by several tenths of magnitude brighter than FU mode pulsators. A
few objects are located by about 1.5--2~mag below the main strip. These
stars, marked by open circles in Figs.~1--4, are almost certainly
Population~II Cepheids. Asterisk and filled squares mark the position of
variable 13682 (V39), 7647 and 13709, respectively, already described in
Section~4.

\setcounter{table}{1}
\MakeTable{crrrr}{7cm}{Best least square fit parameters of the PL
relations:
$M=A\cdot(\log P - 1) + B$}
{
\hline
\hline
\noalign{\vskip2pt}
\multicolumn{5}{c}{IC1613 -- Fundamental Mode Cepheids}\\
\noalign{\vskip2pt}
\hline
\hline
\noalign{\vskip2pt}
Band        &\multicolumn{1}{c}{$A$} &\multicolumn{1}{c}{$B$}& 
\multicolumn{1}{c}{$N$}&\multicolumn{1}{c}{$\sigma$}\\
\noalign{\vskip2pt}
\hline
\noalign{\vskip2pt}
$V$         &$-2.756$      & 20.389      & 64          & 0.193\\
            & 0.095        &  0.041      &             &      \\
$I$         &$-2.946$      & 19.633      & 64          & 0.150\\
            & 0.075        &  0.031      &             &      \\
$W_I$       &$-3.256$      & 18.500      & 56          & 0.111\\
            & 0.057        &  0.024      &             &      \\
\hline
\hline
}

\begin{figure}[p]
\hglue-6mm
\includegraphics[bb=25 50 520 420,width=13cm]{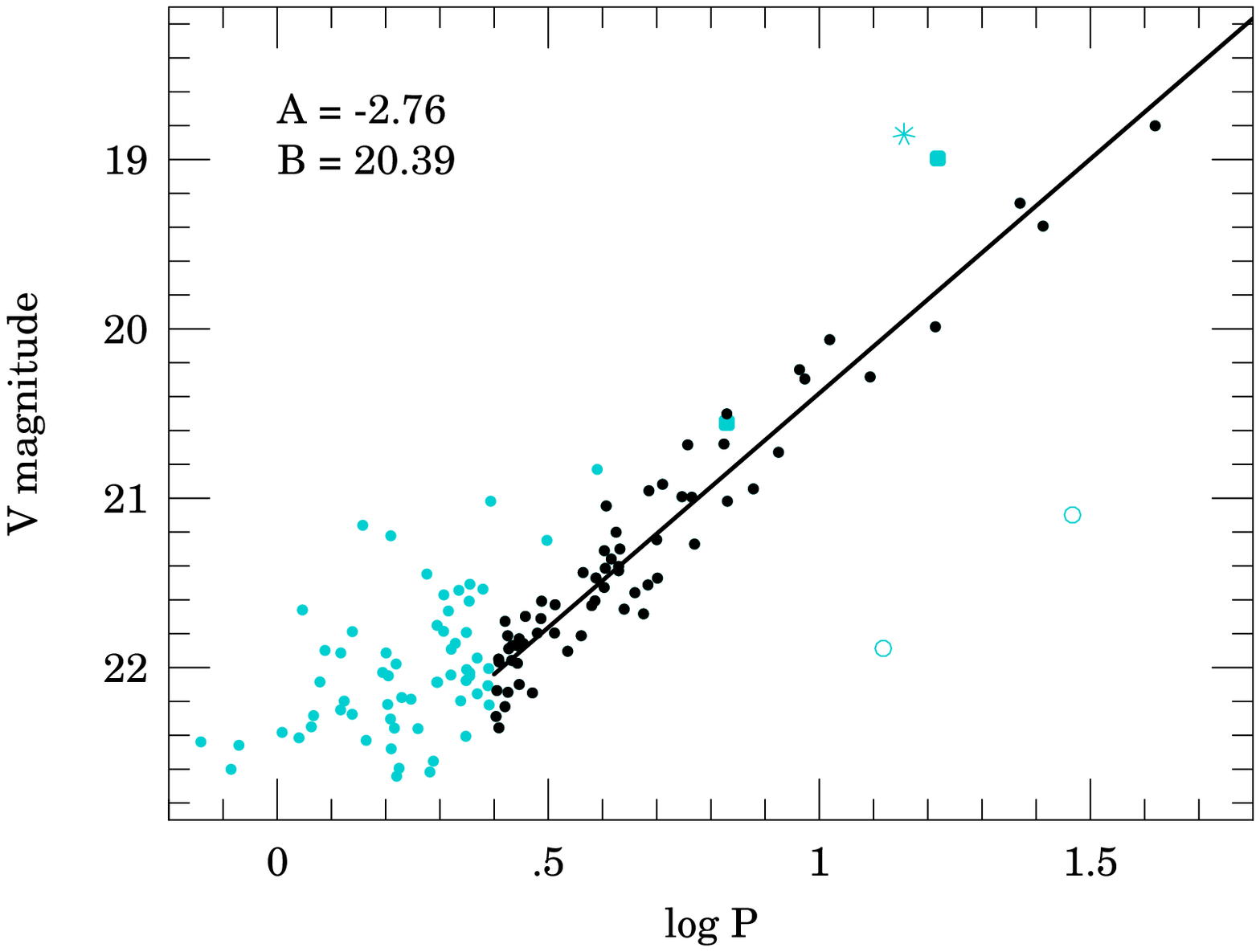}
\vskip3pt
\FigCap{{\it V}-band Period-Luminosity relation for FU mode classical
Cepheids in IC1613. Black dots indicate Cepheids used for the final fit of
PL relation. Gray dots mark Cepheids with period shorter than $\log
P = 0.4$ and objects removed during the iterative fitting of the PL
relation. The remaining symbols are as in Fig.~1.}
\hglue-6mm
\includegraphics[bb=25 50 520 420,width=13cm]{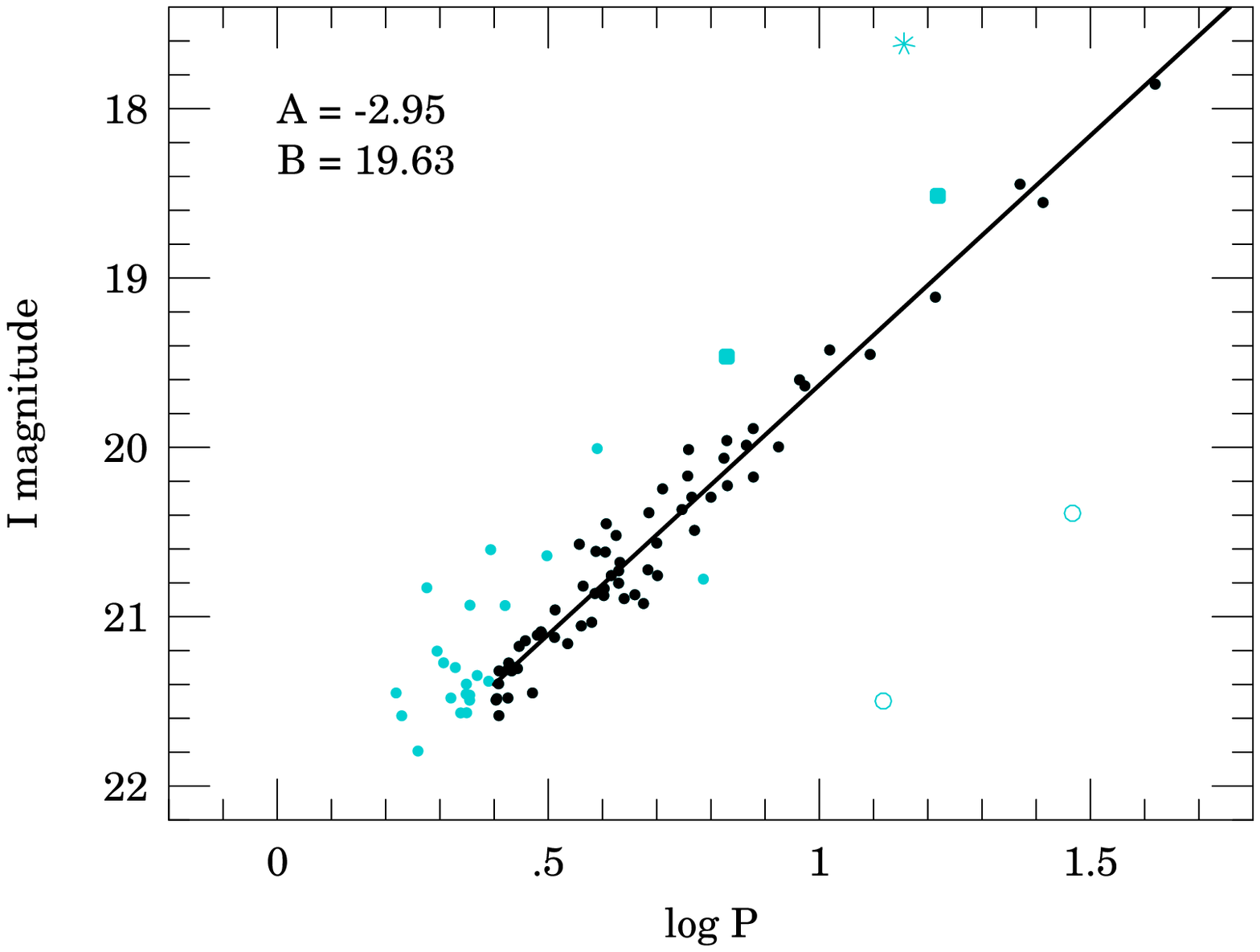}
\vskip3pt
\FigCap{Same as Fig.~2 for the {\it I}-band.}
\end{figure}

\begin{figure}[htb]
\hglue-6mm
\includegraphics[bb=25 50 520 420,width=13cm]{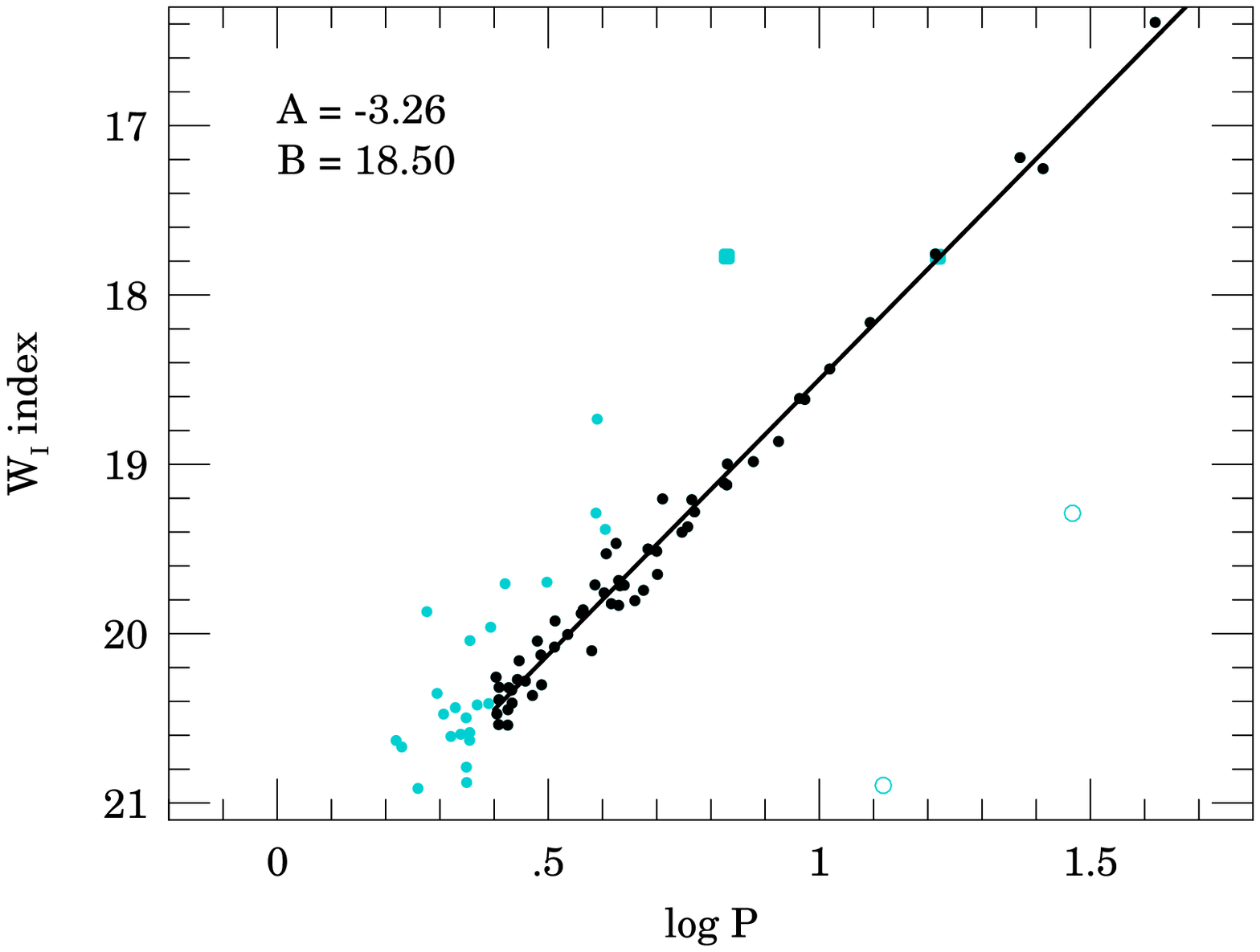}
\vskip3pt
\FigCap{Same as Fig.~2 for the $W_I$ index.}
\end{figure}

To determine the PL relations for the FU mode Classical Cepheids in
IC1613 we followed the same procedure as in Udalski \etal (1999) for
the LMC and SMC. First, we removed the stars classified as likely
Population~II objects. Then, we limited our sample of Cepheids to
objects with $\log P > 0.4$. There are two reasons for such a
limitation. First, in this way we exclude practically all FO Cepheids
contaminating the diagrams at shorter periods. It would be difficult to
separate them from FU mode objects using standard light curve analysis
because of relatively noisy light curves of such faint objects. Secondly,
the number of FU Cepheids with periods shorter than the above limit
varies from object to object. For instance, they are very rare
in the LMC and very numerous in the SMC (Udalski \etal 1999). What
worse, the slope of the PL relation in the SMC for periods shorter than
2~days is steeper (Bauer \etal 1999, Udalski \etal 1999). Therefore to
avoid biases and to have homogeneous material  for comparisons with
Magellanic Cloud Cepheids we further considered  only objects with
periods $\log P > 0.4$.

Next, we determined the PL relation in each of the bands. We fitted
linear relations in the form:
$$ M=A\cdot(\log P - 1)+B \eqno{(2)} $$ 
We applied the same iterative fitting procedure with $2.5\sigma$
clipping algorithm as for the MC Cepheids. In this manner we removed
possible FO pulsators in our sample and/or obvious outliers.

Table~2 lists coefficients of the final PL relations in IC1613 for the
{\it V}, {\it I} and $W_I$ index. The relations  are also plotted as
thick lines in Figs.~2--4. Cepheids used to the final fit are marked by
black dots in these figures. The standard deviation of the fit
decreases from  {\it V}, through {\it I} to the $W_I$ index similar as
in the LMC and SMC and it is in between of that found in the LMC and
SMC.

\Section{Discussion}

Relatively large sample of Cepheids detected in IC1613 makes it possible
to study in detail the PL relations in low metallicity environment of
${\rm [Fe/H]}\approx-1.0$~dex (Dolphin \etal 2001). With so low
metallicity, IC1613 is a crucial object for testing the universality of
the PL relations and accuracy of the Cepheid distance scale. Cepheids
from two remaining objects containing well studied samples of these
variable stars, the LMC and SMC (Udalski \etal 1999), have the mean
metallicity equal to ${\rm [Fe/H]}=-0.3$~dex, and $-0.7$~dex,
respectively (Luck \etal 1998). Therefore comparison of PL relations in
these three galaxies can provide basic information on how PL relations
depend on metallicity of the environment the Cepheid comes from.

\Subsection{Dependence of the PL Relation on Metallicity}

Udalski \etal (1999) presented comparison of the PL relations determined
for large samples of FU mode Cepheids in the LMC and SMC. To within
uncertainties of determination, the slopes of the PL relations   for
{\it V}, {\it I} and $W_I$ index turned out to be  consistent in these
galaxies. Because of much smaller scatter and better population of the
FU mode PL relations in the LMC, their slopes were adopted as universal.
The numbers presented in Udalski \etal (1999) were slightly corrected in
April 2000, due to small recalibration of the OGLE-II photometry and are
equal to $A^{MC}_V=-2.775$, $A^{MC}_I=-2.977$ and $A^{MC}_{W_I}=-3.300$
(Udalski 2000b). Comparing these figures with the IC1613 slopes
presented in Table~2 one immediately finds the striking agreement for
all the presented bands. This result definitively ends the dispute on
the possibility of shallower slopes of PL relations in IC1613 started
almost three decades ago. The main conclusion from this comparison is
that the slopes of the PL relations presented in Udalski (2000b) are
indeed universal and do not depend on metallicity.

To study whether the zero points of the PL relations depend on
metallicity one needs the reference brightness to which the observed
magnitudes of Cepheids could be referenced. Udalski (2000b) proposed the
other major standard candles, namely RR Lyr, red clump and tip of the
red giant branch (TRGB) stars as possible brightness references and
showed that the differences of magnitudes of all major stellar standard
candles are very consistent in the LMC and SMC. The TRGB stars seem to
be the best for testing the stability of zero points of Cepheid  PL
relations. First, they are of brightness similar to Cepheids. Secondly,
they are usually very numerous and the determination of the TRGB
magnitude is very precise (see for instance the TRGB of the LMC and SMC
-- Udalski 2000b). Finally, it is generally accepted that the TRGB {\it
I}-band magnitude is practically constant to better than a few
hundredths of magnitude  for a wide range of  metallicity (Bellazzini,
Ferraro and Pancino 2001). Also, the old TRGB stars belong to completely
different population than young Cepheids so that any correlation of
their luminosities seems to be extremely unlikely.

The difference of the {\it I}-band magnitudes of TRGB stars and Cepheids
with $P=10$~days for the LMC and SMC was determined by Udalski (2000b)
and is equal to: $0.71\pm0.03$~mag and $0.66\pm0.04$~mag for the LMC and
SMC, respectively. It is worth stressing at this point that such a
determination is fully differential and therefore  can be done very
precisely, as the majority of possible  systematic  errors like
uncertainty of the zero point of photometry, uncertainty of interstellar
extinction  etc. cancel out.

There is, however, a possibility that in the case of IC1613 the
magnitudes of Cepheids can be systematically affected by blending
effect. Mochejska \etal (2000, 2001), who first pointed out  this
effect, found that in M31 and M33 galaxies, located at roughly the same
distance as IC1613, some fraction of ground-based detected Cepheids is
blended with close physical/optical companions resolved  on  much higher
resolution HST images. This may lead to a systematic overestimate of the
brightness of Cepheids up to 0.15~mag for these galaxies when measured
from the ground.

Certainly results for M31 and M33, large spiral galaxies, have no direct
correspondence to the possible effect of blending in such low
luminosity, dwarf galaxy as IC1613. In general, blending for a
particular object might be hard to estimate because it may depend on many
local conditions etc. To have some information on the possible magnitude
of this effect in the case of IC1613 we performed two tests. First we
used simulation of Stanek and Udalski (1999) who analyzed how the
blending effect would affect the LMC Cepheids when observed at larger
distances, \ie with worse resolution. LMC is certainly more suitable for
such a test than spiral galaxies M31 and M33 as both the LMC and IC1613
belong to the same class of galaxies. Still, this test can provide only
an upper limit because of much larger brightness, star density etc. of
the LMC bar than the faint dwarf IC1613.

Simple calculation yields that at the distance of IC1613 of about 700
kpc, the ground-based resolution of $1\arcs$ corresponds to the LMC
Cepheids as seen by the HST (resolution of 0\zdot\arcs1) from 7 Mpc.
Fig.~4 from Stanek and Udalski (1999) implies that at this distance the
blending would affect LMC Cepheids at the 0.03--0.04~mag level only. It
seems reasonable then to assume that in the case of IC1613 it will be
considerably smaller.

We also looked at HST images of the center of IC1613 retrieved from the
Hubble Data Archive, originally obtained by Cole \etal (1999). We
limited ourselves to Cepheids used for determination of our PL
relations, \ie those with $\log P > 0.4$. Additionally, we also checked
whether blending similarly affects other group of stars, \ie the TRGB
stars. We selected a subsample of TRGB stars limiting them to $\pm
0.05$~mag around the TRGB mean magnitude (see below). The number of
Cepheids and TRGB stars in the field of Cole \etal (1999) is small and
similar: 10 and 13 objects, respectively. We visually inspected the HST
images looking for blending of identified stars. We found significant
blending in the case of two Cepheids and five TRGB stars.  Although the
samples are small we may conclude that both groups of stars seem to be
similarly affected in the case of IC1613. Because we will further
compare magnitudes of both groups of stars it is safe to conclude that
the possible blending cancels in the first approximation and negligibly
affects our comparison of standard candles in IC1613 presented below. It
will also marginally affect  the slopes of the above determined PL
relations. Blending is supposed to affect more the shorter period
(fainter) Cepheids than the longer period ones. Even if this difference
amounts to $0.05$~mag, the slopes of the PL relation would be just by
similar amount steeper, still in excellent agreement with the slopes of
the Magellanic Cloud Cepheid PL relations. Thus, it seems that, apart
from the Magellanic Clouds, IC1613 might be an unique object for
studying properties of Cepheids not only because of low reddening and
low metallicity but also because possible systematic errors like
blending are in this case practically negligible contrary to, for
instance, M31 or M33 galaxies (Mochejska \etal 2000, 2001).

\begin{figure}[htb]
\hglue-6mm
\includegraphics[bb=25 50 520 420,width=13cm]{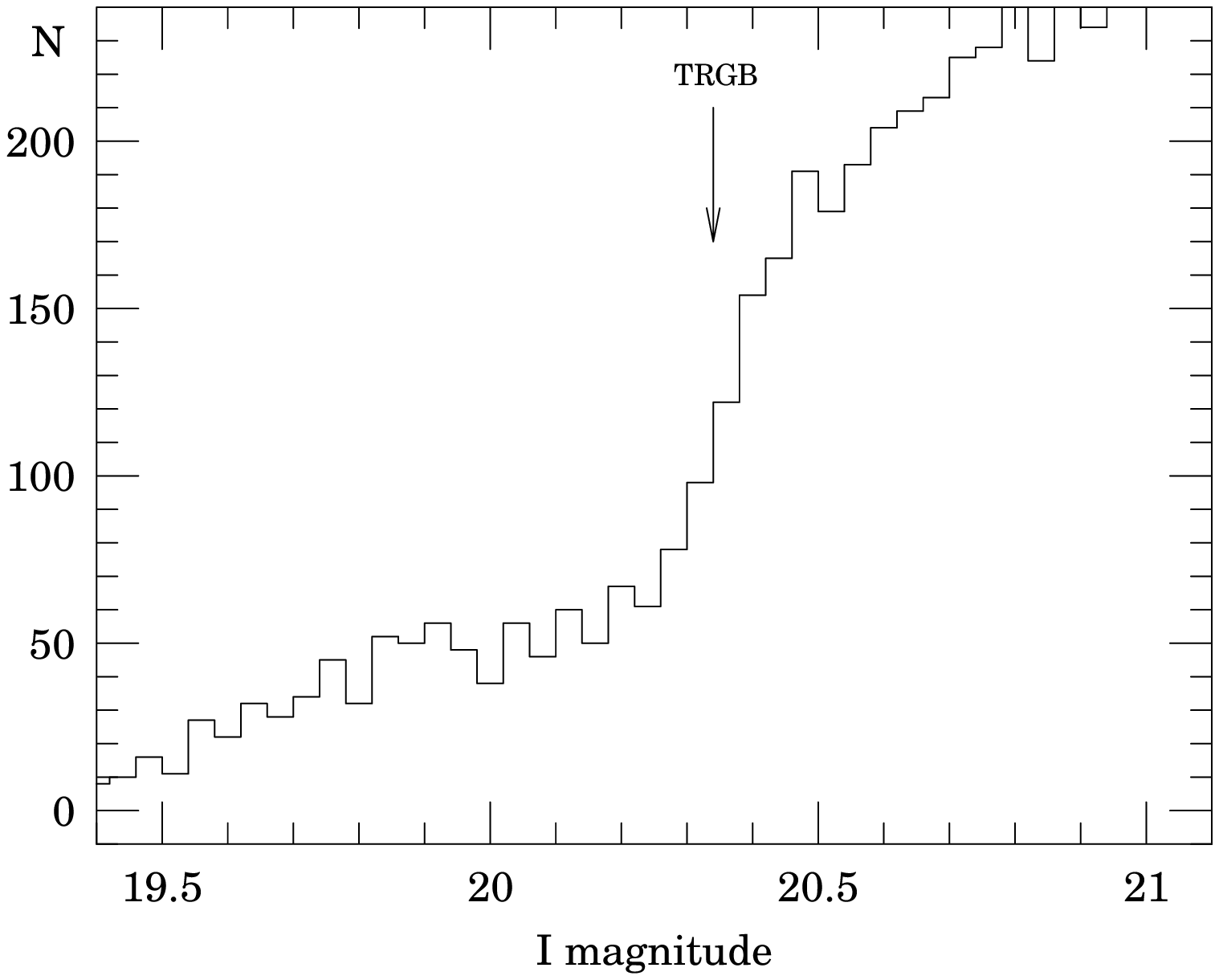}
\vskip3pt
\FigCap{Histogram of luminosity of the upper part of the red
giant branch in IC1613. Bins are 0.04~mag wide. Arrow marks the TRGB
magnitude.}
\end{figure}

CMD diagram of IC1613, presented in Fig.~1, shows that the upper red
giant branch is very well populated in this galaxy and the TRGB is
easily distinguishable. To determine its magnitude we selected the  red
giant branch stars from the CMD region limited by dotted line in Fig.~1.
Then, we constructed the luminosity function histogram with 0.04~mag
bins presented in Fig.~5. The location of the TRGB is clearly seen in
this figure and it is marked by the arrow. The TRGB magnitude in IC1613
is equal to: ${\langle I^{\rm TRGB}\rangle=20.34\pm0.02}$~mag. It is
worth noticing that this value is in very good agreement with other
recent determinations of the TRGB magnitude (see Section~6.2) for IC1613
(Dolphin \etal 2001).

\begin{figure}[htb]
\hglue-6mm
\includegraphics[bb=25 145 520 520,width=13cm]{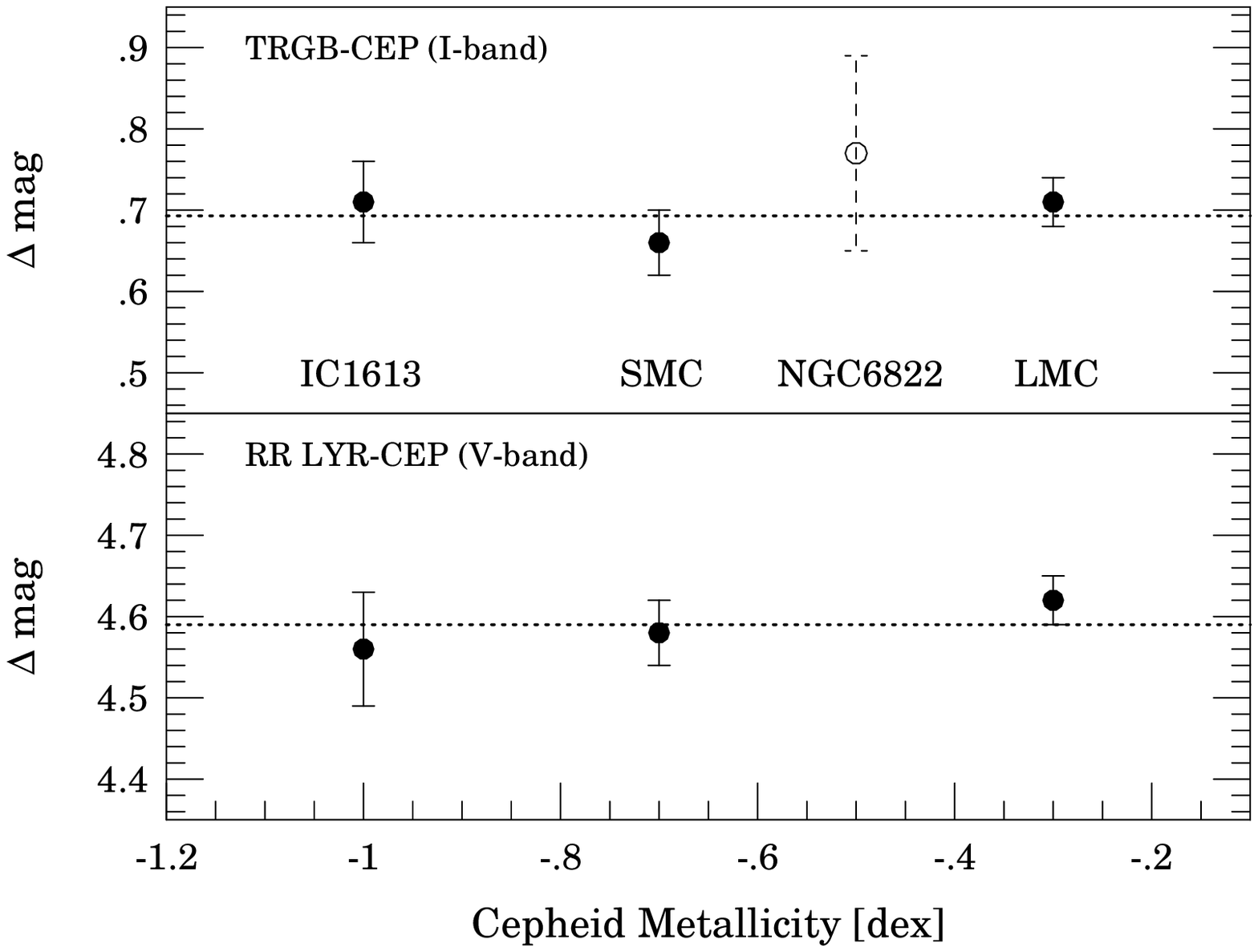}
\vskip3pt
\FigCap{Comparison of brightness of $P=10$~days Cepheids with brightness
of TRGB stars (upper panel) and RR~Lyr stars (lower panel).}
\end{figure}

The {\it I}-band PL relation yields the brightness of the $P=10$~days
Cepheid in IC1613 of ${\langle I^{\rm C}\rangle=19.63\pm0.04}$~mag.
Thus, the difference of magnitudes of the TRGB and   $P=10$~days Cepheid
in IC1613 is equal to $0.71\pm0.05$~mag. Table~3 summarizes comparison
of the {\it I}-band TRGB and Cepheid magnitudes in the LMC, SMC and
IC1613. Upper panel of Fig.~6 visualizes these results. Looking at Fig.~6
one can conclude that in all three objects distributed over wide range
of Cepheid metallicities, the mean {\it I}-band brightness of Cepheids
relative to the reference source, TRGB stars, is constant to
$\pm0.03$~mag.

It should be noted here that at least one additional object could be in
principle added to Fig.~6. Another Local Group dwarf galaxy, NGC6288, is
known to contain both Cepheids and TRGB stars. With metallicity of ${\rm
[Fe/H]}\approx-0.5$~dex (Venn \etal 2001) it could provide another point
in between the LMC and SMC. Unfortunately, similarly to IC1613, NGC6822
still awaits for extensive, standard band search for its Cepheids and
determination of precise PL relations. Nevertheless, the rough estimate
of the difference of brightness of TRGB stars and Cepheids can be
obtained from the data of Gallart, Aparicio and V\'{\i}lchez (1996).
They determine the {\it I}-band magnitude of TRGB stars in NGC6822 to be
equal to ${\langle I^{\rm TRGB}\rangle=19.8\pm0.1}$~mag. They also list
the mean {\it I}-band magnitudes for six Cepheids from this galaxy. 
Fitting the universal PL relation ($A_I=-2.98$) to these data gives the
brightness of the $P=10$ days Cepheid equal to ${\langle I^{\rm
C}\rangle=19.03\pm0.07}$~mag leading to the difference of
$0.77\pm0.12$~mag. It is very encouraging that this value is in very
good agreement with the MC and IC1613 data confirming the observed 
behavior of Cepheids in environments of different metallicity.
Nevertheless, we treat this value as very preliminary and mark  in
Fig.~6 the point for NGC6822 as an open circle with dashed line error
bars until a more precise value for NGC6822 is available. One should
also remember that a few more nearby galaxies are known to contain
Cepheids (\eg WLM, NGC3109, NGC300). They may also provide additional
points to Fig.~6 in the future when accurate and based on many Cepheids
standard band PL relations, precise TRGB magnitudes and accurate
spectroscopic metallicities are  determined.

\MakeTable{cccc}{12.5cm}{Comparison of brightness of Cepheids with TRGB
and RR~Lyr stars}
{\hline
\noalign{\vskip4pt}
Galaxy:& IC1613 & SMC & LMC\\
Cepheid [Fe/H]:& $-1.0$~dex& $-0.7$~dex & $-0.3$~dex\\
\noalign{\vskip3pt}
\hline
\noalign{\vskip4pt}
TRGB -- CEPHEIDS & & & \\
\noalign{\vskip3pt}
$ \langle I^{\rm TRGB}\rangle-\langle I^{\rm C}\rangle$
& $0.71\pm0.05$ & $0.66\pm0.04$ & $0.71\pm0.03$ \\
\noalign{\vskip4pt}
\hline
\noalign{\vskip4pt}
RR~LYR -- CEPHEIDS & & & \\
\noalign{\vskip3pt}
$ \langle V^{\rm RR}\rangle_{{\rm [Fe/H]}_{\rm LMC}}-\langle V^{\rm
C}\rangle$ & $4.56\pm0.07$ & $4.58\pm0.04$ & $4.62\pm0.03$ \\
\noalign{\vskip4pt}
\hline
}

The comparison of the mean magnitudes of Cepheids and TRGB stars implies
no metallicity dependence of the {\it I}-band zero point. Although it is
unlikely to expect significantly different behavior of Cepheids in
shorter wavelength bands, this cannot be in principle excluded.
Fortunately, IC1613 also contains RR Lyr stars which {\it V}-band
magnitude is believed to be a good standard candle. However, contrary to
the TRGB stars magnitude, the RR Lyr brightness is known to be dependent
on metallicity. The slope of this dependence is usually assumed to be
about 0.15--0.20~mag/dex with RR Lyr being brighter in lower metallicity
environment. 

13 RR~Lyr stars in IC1613 were detected by Dolphin \etal (2001) on deep
HST images. They provide the mean {\it V}-band brightness of this sample
to be equal to ${\langle V^{\rm RR}\rangle=25.00\pm0.04}$~mag. They also
determined the mean metallicity of this sample to be ${\rm
[Fe/H]}=-1.3\pm0.2$~dex. 

Udalski (2000b) finds the mean differences of the {\it V}-band
brightness of RR Lyr stars, converted to the mean metallicity of the LMC
RR~Lyr stars (${\rm [Fe/H]}=-1.6$~dex) and $P=10$~days Cepheids in the
LMC and SMC to be equal to $4.62\pm0.03$~mag and $4.58\pm0.04$~mag,
respectively. Unfortunately, similar comparison for IC1613 cannot be
done fully differentially in our case, contrary to the  Magellanic Cloud
or {\it I}-band comparisons, because some shift between the zero points
of Dolphin \etal (2001) and our {\it V}-band photometries is possible.
Although, in the case of the {\it I}-band TRGB magnitudes the agreement
of Dolphin  \etal (2001) and our determinations is excellent we assume
additional uncertainty of $\pm0.02$~mag for the possible difference of
{\it V}-band photometries.

Before we compare the {\it V}-band magnitudes of RR Lyr and Cepheids we
must first convert the former to the metallicity of the LMC. If the
IC1613 RR Lyr stars were of the LMC metallicity they would have to be by
about 0.05~mag brighter, \ie ${\langle V^{\rm RR}\rangle_{{\rm [Fe/H]}_{\rm
LMC}}=24.95\pm0.05}$~mag. With the mean {\it V}-band magnitude of
$P=10$~days Cepheids equal to ${\langle V^{\rm
C}\rangle=20.39\pm0.04}$~mag this leads to the difference equal to
$4.56\pm0.07$~mag. In the error budget we additionally included
$0.02$~mag uncertainty of ground-based Cepheid magnitude for possible
blending effect. Results of this comparison are also listed in Table~3
and plotted in the lower panel of Fig.~6.

Fig.~6 clearly indicates that brightness of Cepheids in respect to
RR~Lyr stars in the {\it V}-band behaves basically identical to our
previous comparison with TRGB stars in the {\it I}-band. Again, the
brightness of Cepheids is to $\pm0.04$~mag constant in all objects
covering wide range of Cepheid metallicities implying no dependence of
the {\it V}-band zero point of the PL relation on metallicity.

Summarizing,  we can conclude that both {\it VI} and $W_I$ slopes and
also {\it VI} zero points of the Cepheid PL relation are independent of
metallicity in the wide range of $-1.0<{\rm [Fe/H]} < -0.3$~dex. Our
tests clearly  indicate that the PL relations are universal in this
range, so the distance scale based on Cepheids is very reliable and
Cepheids are very good standard candles.  Of course, we cannot exclude
at this moment that some dependence on metallicity occurs outside this
range, for instance, for more metal rich object of Galactic metallicity
of ${\rm [Fe/H]}\approx 0.0$~dex, or higher. However, analysis of this
range will probably have to be postponed until the future space missions
(GAIA, FAME, SIM)  provide precise absolute magnitudes of large samples
of Galactic Cepheids, RR~Lyr and TRGB stars.

\Subsection{Distance to IC1613}

Beside Cepheids, TRGB and RR~Lyr stars, IC1613 also contains large
number of red clump stars. These stars are  also believed to be a very
good standard candle, the only one which can be directly calibrated with
Hipparcos parallaxes (Paczy{\'n}ski and Stanek 1998, Udalski 2000a). 

While the observational tests show that the mean {\it I}-band magnitude
of red clump stars is only slightly dependent on metallicity of
environment (the slope of 0.14~mag/dex with red clump stars being
brighter in metal poor objects) and practically independent of the age
for intermediate age (2--10~Gyr) objects (Udalski 2000ab, Bersier 2000),
theoretical modeling predicts its complicated behavior questioning
usability of red clump stars as a standard candle (Girardi and Salaris
2001).

The problem is very important because the red clump stars have very
reliable calibration based on Hipparcos measurements of nearby stars
(Udalski 2000a). Therefore they might be crucial for establishing the
precise zero point of the distance scale in the Universe.

Bersier (2000) and Udalski (2000b) presented comparison of the mean {\it
I}-band magnitudes of red clump stars with TRGB stars. This comparison,
fully differential and largely free from systematic errors, showed that
after correcting for the above mentioned slight dependence of the red
clump brightness on metallicity, the difference $ \langle I^{\rm
RC}\rangle_{{\rm [Fe/H]}_{\rm LMC}}-\langle I^{\rm TRGB}\rangle$ is
equal to $3.62\pm0.05$~mag for ten galaxies possessing red clump stars
of intermediate age and believed to have  wide age distribution,
different star formation histories etc. (Table~5 in Udalski 2000b). This
result clearly indicates that the dependence of the mean {\it I}-band
magnitude of red clump stars on age in the range of 2--10~Gyr is
marginal.

IC1613 offers another opportunity for testing this assumption. Dolphin
\etal (2001) provide the mean {\it I}-band magnitudes for both the red
clump and TRGB stars obtained from deep HST images. Again the comparison
can be done fully differentially because both kind of stars were observed
simultaneously. Dolphin \etal (2001) provide data for their field and
re-reduced data for  another field in IC1613 observed by Cole \etal
(1999): $ \langle I^{\rm RC}\rangle=23.90\pm0.01$ and
$23.86\pm0.01$~mag, respectively, and $\langle I^{\rm
TRGB}\rangle=20.40\pm0.09$ and $20.35\pm0.07$~mag, respectively. 

Before making the comparison one has to correct the brightness of red
clump stars for the difference of metallicity LMC -- IC1613: $-0.5$~dex
\vs $-1.3$~dex (Cole \etal 1999, Dolphin \etal 2001) by applying
correction of $+0.11$~mag (Eq.~14, Udalski 2000b). The final difference
$\langle I^{\rm RC}\rangle_{{\rm [Fe/H]}_{\rm LMC}}-\langle I^{\rm
TRGB}\rangle$, very similar for both Dolphin \etal (2001) fields, is
equal to $3.62\pm0.07$~mag. The main component of the error in this
determination is the cited error of the TRGB magnitude. In practice, it
is very likely that the real determination error is at least twice
smaller -- our ground based observations give similar value of TRGB
magnitude with much smaller uncertainty.

The comparison of the {\it I}-band magnitudes of the red clump and TRGB
stars in IC1613 yields identical result as for the remaining ten
galaxies from Table~5 of Udalski (2000b). It is worth noticing that
according to Cole \etal (1999) the age of the red clump stars in IC1613
is about 7~Gyr, rather on the older side of the intermediate age range. 
Nevertheless, the difference of magnitudes of the red clump and TRGB
stars is identical as in objects possessing much younger population of
red clump stars like the LMC (2--3~Gyr). Thus, IC1613 is eleventh object
confirming that the red clump {\it I}-band magnitude is independent of
age and only slightly dependent on metallicity. Therefore, there is no
empirical evidence to question the Hipparcos calibration of red clump
stars for establishing the zero point of the distance scale.

The most likely absolute calibrations of the distance scale for the main
stellar standard candles: Cepheids, RR Lyr, TRGB and red clump stars are
given in Eqs.~9--14 of Udalski (2000b). They give consistent distances
to the LMC, SMC and Carina dwarf galaxy for all the observed there
stellar distance indicators.  These calibrations are in agreement with
differences of magnitudes between standard candles observed in
particular objects. They assume only slight dependence of magnitudes on
metallicity in the case of RR Lyr and red clump stars (well established
empirically) and no other population effects.  The calibrations are
based on the only direct calibration of standard candles -- Hipparcos
calibration of red clump stars, but are also in agreement with other
less direct calibrations of the remaining distance indicators. 

\MakeTable{ccc}{12.5cm}{Distance moduli to IC1613}
{\hline
\noalign{\vskip3pt}
STANDARD CANDLE & $m_0$ & $(m-M)_{\rm IC1613}$\\
\noalign{\vskip2pt}
\hline
\noalign{\vskip5pt}
CEPHEIDS ({\it V}) & $20.31\pm0.05$ & $24.23\pm0.07$\\
\noalign{\vskip3pt}
CEPHEIDS ({\it I}) & $19.58\pm0.04$ & $24.19\pm0.07$\\
\noalign{\vskip3pt}
CEPHEIDS ($W_I$)   & $18.50\pm0.03$ & $24.17\pm0.07$\\
\noalign{\vskip3pt}
RR~LYR ({\it V})& $24.92\pm0.04$ & $24.18\pm0.07$\\
\noalign{\vskip3pt}
TRGB ({\it I})& $20.29\pm0.02$ & $24.20\pm0.07$\\
\noalign{\vskip3pt}
RED CLUMP ({\it I}) & $23.83\pm0.02$ & $24.23\pm0.06$\\
\noalign{\vskip3pt}
\hline}

In Table~4 we present the distance determinations to IC1613 resulting
from four stellar standard candles which photometric data are presented
in this paper. The observed magnitudes were corrected for interstellar
extinction using the values listed in Section~4 (column $m_0$ in
Table~4). For Cepheids three determination were possible -- for the {\it
V}, {\it I} and $W_I$ index.

As can be seen from Table~4, the resulting  distance moduli from major
stellar distance indicators in IC1613 are very consistent similarly to
the LMC, SMC and Carina dwarf galaxy determinations (Udalski 2000b). The
distance modulus to IC1613 is equal to $(m-M)_{\rm IC1613}=24.20$~mag
with the standard deviation from six determinations of only
$\pm0.02$~mag. The systematic uncertainty, however, can be of the order
of 0.07~mag resulting from uncertainty of Udalski's (2000b)
calibrations, error of determination of the mean magnitudes and
additional possible blending effect in IC1613 in the case of
ground-based measurements (estimated at $+0.02$~mag). The distance
modulus of IC1613 corresponds to the distance of  $690\pm20$~kpc.

The photometric data of IC1613 presented in this paper are available
in the electronic form from the OGLE archive: 
\begin{center}
{\it http://www.astrouw.edu.pl/\~{}ogle} \\
{\it ftp://sirius.astrouw.edu.pl/ogle/ogle2/var\_stars/ic1613}\\
\end{center}
or its US mirror
\begin{center}
{\it http://bulge.princeton.edu/\~{}ogle}\\
{\it ftp://bulge.princeton.edu/ogle/ogle2/var\_stars/ic1613}\\
\end{center}

\Acknow{We would like to thank Drs B. Paczy{\'n}ski and K.Z.\ Stanek for
comments and remarks on the paper. We also thank Ms.\ B.\ Mochejska
for her help in retrieving the HST images of IC1613. This work was
partially based on observations with the NASA/ESA Hubble Space Telescope,
obtained from the data Archive at the Space Telescope Science Institute,
which is operated by the Association of Universities for Research in
Astronomy, Inc. under NASA contract No.\ NAS5--26555. The paper was
partly supported by the  Polish KBN grant 2P03D01418 to M.\ Kubiak.
Partial support for the OGLE project was provided with the NSF  grant
AST-9820314 to B.~Paczy\'nski.}

\end{document}